\documentclass[12]{article}
\usepackage{amssymb}
\usepackage{graphics} 
\usepackage[lmargin=1.7in,rmargin=1.5in,tmargin=1.4in,bmargin=1.2in]{geometry}
\begin{document}
\mbox{}\\[1mm]
\begin{center}
{\LARGE \bf A quantum mechanical model for the\\[1mm] relationship between stock price and\\[2mm] stock ownership}\\[5mm]
{Liviu-Adrian Cotfas}\\[5mm]
{\it Faculty of Economic Cybernetics, Statistics and Informatics,\\ Academy of Economic Studies,  Bucharest, Romania\\ E-mail:   \texttt{lcotfas@gmail.com}}
\end{center}
\bigskip
{\bf Abstract}. The trade of a fixed stock can be regarded as the basic process that measures its momentary price.
The stock price is exactly known only at the time of sale when the stock is between traders, that is, only in the case when the owner is unknown. We show that the stock price can be better described by a function indicating at any moment of time the probabilities for the possible values of price if a transaction takes place. This more general description contains partial information on the stock price, but it also contains partial information on the stock owner. By following the analogy with quantum mechanics, we assume that the time evolution of the function describing the stock price can be described by a Schr\" odinger type equation.

\bigskip

\section{Introduction} \label{introd}

The stock price can only be determined at the time of sale when it is between traders. 
We can never simultaneously know exactly  both the ownership of a stock and its price:
\begin{itemize}
\item
If we know the owner then we have no information on the stock price.
\item
If we know the stock price then we have no information on the owner.
\end{itemize}
In quantum mechanics \cite{Messiah} we have a very similar situation:
\begin{itemize}
\item
If we know the position of a particle then  we have no information on its momentum.
\item
If we know the momentum of a particle then we have no information on its position.
\end{itemize}

In quantum mechanics, the two situations (exactly known position, exactly known momentum) represent only particular states of the considered quantum system. 
The general state, described by a normalized wavefunction $\psi $, generally,  contains partial information on both position and momentum:
\begin{itemize}
\item
$|\psi (x)|^2$ represents the (density of) probability to find the particle at point $x$
\item
 $|F[\psi] (p)|^2$, where
\[
F[\psi] (p)=\frac{1}{\sqrt{2\pi \hbar}}\int {\rm e}^{-{\rm i}px/\hbar }\, \psi (x)\, dx
\]
is the Fourier transform of $\psi $, represents  the (density of) probability for our particle to have a momentum equal to $p$.
\end{itemize}

By following the analogy with quantum mechanics, we propose a more general description for the price of a stock which allows us to distinguish cases in which we have only partial information on the stock price but at the same time partial information on the owner. In the proposed description the cases when the price or owner is exactly known represent just particular cases. 

The mathematical modeling of price dynamics of the financial market is a very complex problem \cite{Baaquie}. We could never take into account all economic and non-economic conditions that have influences to the market. Therefore, we usually consider some very simplified and idealized models, a kind of toy models which mimic certain features of a real stock market \cite{Bagarello,Zhang,Pedram,Delcea}. 
We consider that a finite-dimensional Hilbert space is sufficient for our purpose, and our model is based on the mathematical formalism used in the case of quantum systems with finite-dimensional Hilbert space \cite{Vourdas}. This formalism is essentially based on the finite Fourier transform.

\section{Finite Fourier transform}

Let $N>0$ be a fixed integer number. The space $\mathcal{H}$ of all the complex functions 
\[
\varphi  :\{0,1,..., N\!-\!1\}\longrightarrow \mathbb{C}
\]
can be identified with the Hilbert space $\mathbb{C}^N$ by using the one-to-one mapping
\[
\mathcal{H}\longrightarrow \mathbb{C}^N:\ \varphi \mapsto (\varphi (0), \varphi (1),...,\varphi (N\!-\!1))
\]
if we define in $\mathcal{H}$ the addition of two functions 
\[
(\varphi +\psi )(n)=\varphi (n)+\psi (n)
\]
the multiplication of a function by a complex number
\[
(\lambda \, \varphi )(n)=\lambda \, \varphi (n)
\]
and the scalar product of  two functions
\[
\langle \varphi ,\psi \rangle =\sum _{n=0}^{N-1}\overline{\varphi (n)}\, \psi (n).
\]
The number $||\varphi ||=\sqrt {\langle \varphi ,\varphi \rangle }$ represents the norm of $\varphi $, and the function
\[
\Phi  :\{0,1,..., N\!-\!1\}\longrightarrow \mathbb{C}, \qquad \Phi (n)=\frac{\varphi (n)}{||\varphi ||}
\]
satisfying the relation 
\[
|\Phi (0)|^2+|\Phi (1)|^2+...+|\Phi (N\!-\!1)|^2=1
\]
represents the normalized function corresponding to $\varphi $.

The transformation, called the finite Fourier transformation,
\[
\mathcal{H}\longrightarrow \mathcal{H}:\varphi \mapsto \mathcal{F}[\varphi ],\qquad \mathcal{F}[\varphi ](k)=\frac{1}{\sqrt{N}}\sum _{n=0}^{N-1}{\rm e}^{-\frac{2\pi {\rm i}}{N}kn}\, \varphi (n)
\]
 is one-to-one, and the corresponding inverse transformation is 
\[
\mathcal{H}\longrightarrow \mathcal{H}:\varphi \mapsto \mathcal{F}^{-1}[\psi ],\qquad \mathcal{F}^{-1}[\psi ](n)=\frac{1}{\sqrt{N}}\sum _{k=0}^{N-1}{\rm e}^{\frac{2\pi {\rm i}}{N}kn}\, \psi (k).
\]
More than that, it is a unitary transformation: for any $\varphi $ and $\psi $ we have
\[
\langle \mathcal{F}[\varphi ],\mathcal{F}[\psi ]\rangle =\langle \varphi ,\psi \rangle \qquad {\rm and}\qquad || \mathcal{F}[\varphi ]||=||\varphi ||.
\]
{\bf Example1}. Let $m\!\in \!\{0,1,..., N\!-\!1\}$ be a fixed element. The finite Fourier transform of 
\[
\delta _m  :\{0,1,..., N\!-\!1\}\longrightarrow \mathbb{C}, \qquad \delta _m(n)=\left\{
\begin{array}{lll}
1 & {\rm if} & n=m\\
0 & {\rm if} & n\neq m
\end{array} \right.
\]
is the function 
\[
\mathcal{F}[\delta _m ]:\{0,1,..., N\!-\!1\}\longrightarrow \mathbb{C}, \qquad \mathcal{F}[\delta _m ](k)=\frac{1}{\sqrt{N}}\, {\rm e}^{-\frac{2\pi {\rm i}}{N}km}
\]
with property (see Figure 1)
\[
|\mathcal{F}[\delta _m ](0)|^2=|\mathcal{F}[\delta _m ](1)|^2=\cdots =|\mathcal{F}[\delta _m  ](N\!-\!1)|^2=\frac{1}{N}.
\]

\begin{center}
\begin{figure}[h]
\setlength{\unitlength}{2mm}
\begin{picture}(70,20)(0,0)
\put(2,3){\vector(1,0){30}}
\put(37,3){\vector(1,0){30}}

\put(5,0){\vector(0,1){18}}
\put(40,0){\vector(0,1){18}}

\put(    5.00000,   3){\circle*{0.3}}
\put(    6.20000,   3){\circle*{0.3}}
\put(    7.40000,   3){\circle*{0.3}}
\put(    8.60000,   3){\circle*{0.3}}
\put(    9.80000,   3){\circle*{0.3}}
\put(   11.00000,   3){\circle*{0.3}}
\put(   12.20000,  3){\circle*{0.3}}
\put(   13.40000,  15){\circle*{0.3}}
\put(   14.60000,   3){\circle*{0.3}}
\put(   15.80000,   3){\circle*{0.3}}
\put(   17.00000,   3){\circle*{0.3}}
\put(   18.20000,   3){\circle*{0.3}}
\put(   19.40000,   3){\circle*{0.3}}
\put(   20.60000,   3){\circle*{0.3}}
\put(   21.80000,   3){\circle*{0.3}}
\put(   23.00000,   3){\circle*{0.3}}
\put(   24.20000,   3){\circle*{0.3}}
\put(   25.40000,   3.00000){\circle*{0.3}}
\put(   26.60000,   3.00000){\circle*{0.3}}
\put(   27.80000,   3.00000){\circle*{0.3}}
\put(   29.00000,   3.00000){\circle*{0.3}}

\put(   13.40000,3){\line(0,1){   12}}

\put(   40.00000,   3.57){\circle*{0.3}}
\put(   41.20000,   3.57){\circle*{0.3}}
\put(   42.40000,   3.57){\circle*{0.3}}
\put(   43.60000,   3.57){\circle*{0.3}}
\put(   44.80000,   3.57){\circle*{0.3}}
\put(   46.00000,   3.57){\circle*{0.3}}
\put(   47.20000,   3.57){\circle*{0.3}}
\put(   48.40000,   3.57){\circle*{0.3}}
\put(   49.60000,   3.57){\circle*{0.3}}
\put(   50.80000,   3.57){\circle*{0.3}}
\put(   52.00000,   3.57){\circle*{0.3}}
\put(   53.20000,   3.57){\circle*{0.3}}
\put(   54.40000,   3.57){\circle*{0.3}}
\put(   55.60000,  3.57){\circle*{0.3}}
\put(   56.80000,  3.57){\circle*{0.3}}
\put(   58.00000,  3.57){\circle*{0.3}}
\put(   59.20000,   3.57){\circle*{0.3}}
\put(   60.40000,   3.57){\circle*{0.3}}
\put(   61.60000,   3.57){\circle*{0.3}}
\put(   62.80000,   3.57){\circle*{0.3}}
\put(   64.00000,   3.57){\circle*{0.3}}
\put(   40.00000,3){\line(0,1){    0.57}}
\put(   41.20000,3){\line(0,1){   0.57}}
\put(   42.40000,3){\line(0,1){    0.57}}
\put(   43.60000,3){\line(0,1){   0.57}}
\put(   44.80000,3){\line(0,1){   0.57}}
\put(   46.00000,3){\line(0,1){   0.57}}
\put(   47.20000,3){\line(0,1){   0.57}}
\put(   48.40000,3){\line(0,1){    0.57}}
\put(   49.60000,3){\line(0,1){    0.57}}
\put(   50.80000,3){\line(0,1){   0.57}}
\put(   52.00000,3){\line(0,1){   0.57}}
\put(   53.20000,3){\line(0,1){   0.57}}
\put(   54.40000,3){\line(0,1){   0.57}}
\put(   55.60000,3){\line(0,1){   0.57}}
\put(   56.80000,3){\line(0,1){  0.57}}
\put(   58.00000,3){\line(0,1){   0.57}}
\put(   59.20000,3){\line(0,1){   0.57}}
\put(   60.40000,3){\line(0,1){    0.57}}
\put(   61.60000,3){\line(0,1){   0.57}}
\put(   62.80000,3){\line(0,1){   0.57}}
\put(   64.00000,3){\line(0,1){   0.57}}

\put(4.2,1.8){$\scriptscriptstyle{0}$}
\put(28.5,1.8){$\scriptscriptstyle{20}$}
\put(16.5,1.8){$\scriptscriptstyle{10}$}
\put(10.5,1.8){$\scriptscriptstyle{5}$}
\put(22.5,1.8){$\scriptscriptstyle{15}$}

\put(39.2,1.8){$\scriptscriptstyle{0}$}
\put(63.5,1.8){$\scriptscriptstyle{20}$}
\put(51.5,1.8){$\scriptscriptstyle{10}$}
\put(45.5,1.8){$\scriptscriptstyle{5}$}
\put(57.5,1.8){$\scriptscriptstyle{15}$}

\put(31,3.3){$\scriptstyle{n}$}
\put(66,3.3){$\scriptstyle{k}$}

\put(3,8.5){$\scriptscriptstyle{0.5}$}
\put(4.9, 9){\line(1,0){0.2}}
\put(4,14.5){$\scriptscriptstyle{1}$}
\put(4.9, 15){\line(1,0){0.2}}
\put(38,8.5){$\scriptscriptstyle{0.5}$}
\put(39.9, 9){\line(1,0){0.2}}
\put(39,14.5){$\scriptscriptstyle{1}$}
\put(39.9, 15){\line(1,0){0.2}}

\put(5.4,17){$\scriptscriptstyle{|\delta _m (n)|^2}$}
\put(40.4,17){$\scriptscriptstyle{|\mathcal{F}[\delta _m ](k)|^2}$}

\end{picture}
\caption{$|\delta _m (n)|^2$ and $|\mathcal{F}[\delta _m](k)|^2$ in the case when $N\!=\!21$ and $m\!=\!7$.}
\end{figure}
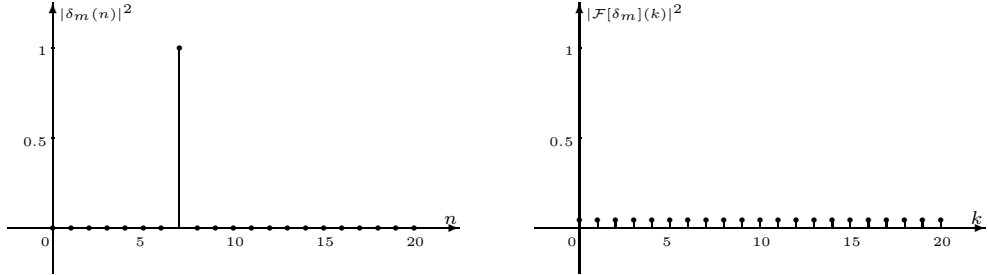
\end{center}

{\bf Example 2}. The Jacobi theta function \cite{Magnus,Vilenkin,Whittaker}
\[
\theta _3(z,\tau )=\sum_{\alpha =-\infty }^\infty {\rm e}^{{\rm i}\pi \tau \alpha ^2}\, {\rm e}^{2\pi {\rm i}\alpha z},\qquad \mathfrak{Im}( \tau )>0
\]
has several remarkable properties among which we mention:
\[
\theta _3(z+m+n\tau ,\tau )={\rm e}^{-{\rm i}\pi \tau n^2}\, {\rm e}^{-2\pi {\rm i}nz}\, \theta _3(z,\tau )
\]
\[
\theta _3(z,{\rm i}\tau )=\frac{1}{\sqrt{\tau }}\, {\rm exp}^{-\frac{\pi z^2}{\tau }}\, \theta _3 \left( \frac{z}{{\rm i}\tau },\frac{\rm i}{\tau }\right)
\]
and \cite{Ruzzi,Mehta}
\[
\theta _3 \left( \frac{k}{N},\frac{{\rm i}\kappa }{N}\right)=\frac{1}{\sqrt{\kappa N}}\, \sum_{n=0}^{N-1}{\rm e}^{-\frac{2\pi {\rm i}}{N}kn}\, \theta _3 \left( \frac{n}{N},\frac{\rm i}{ \kappa N}\right).
\]

For any $\kappa \in (0,\infty )$ the function 
\[
\gamma _\kappa  :\{0,1,..., N\!-\!1\}\longrightarrow \mathbb{C}, \qquad \gamma _\kappa  (n)=\sum _{m=-\infty }^\infty {\rm e}^{-\frac{\kappa \pi }{N}(mN+n)^2}
\]
can be expressed in terms of the Jacobi function $\theta _3$ as
\[
\gamma _\kappa (n)=\frac{1}{\sqrt{\kappa N}}\, \theta _3\left( \frac{n}{N},\frac{{\rm i}}{\kappa N} \right)
\]
and satisfies the relation
\[
\mathcal{F}[\gamma _\kappa ]=\frac{1}{\sqrt{\kappa }}\, \gamma _{\frac{1}{\kappa }}.
\]
Since $||\gamma _{\frac{1}{\kappa }}||=\sqrt{\kappa }\, ||\mathcal{F}[\gamma _\kappa ]||=\sqrt{\kappa }\, ||\gamma _\kappa ||$, the normalized function 
\[
\Upsilon _\kappa :\{0,1,..., N\!-\!1\}\longrightarrow \mathbb{C}, \qquad \Upsilon _\kappa (n)=\frac{\gamma _\kappa (n)}{||\gamma _\kappa ||}
\]
satisfies the relation
\[
\mathcal{F}[\Upsilon _\kappa ]=\Upsilon _{\frac{1}{\kappa }}.
\]

\begin{center}
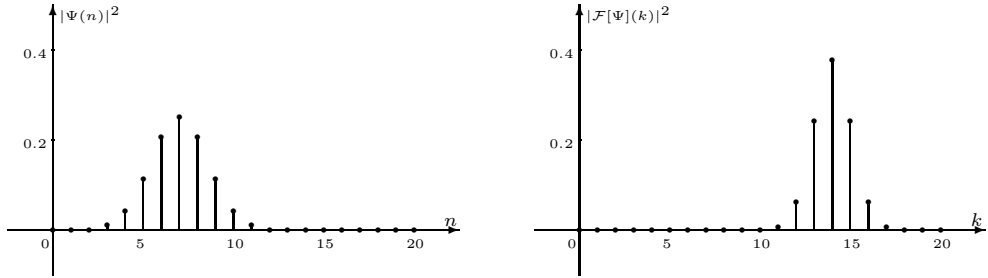
\begin{figure}[h]
\setlength{\unitlength}{2mm}
\begin{picture}(70,20)(0,0)
\put(2,3){\vector(1,0){30}}
\put(37,3){\vector(1,0){30}}

\put(5,0){\vector(0,1){18}}
\put(40,0){\vector(0,1){18}}

\put(    5.00000,   3.00043){\circle*{0.3}}
\put(    6.20000,   3.00575){\circle*{0.3}}
\put(    7.40000,   3.05162){\circle*{0.3}}
\put(    8.60000,   3.31078){\circle*{0.3}}
\put(    9.80000,   4.25556){\circle*{0.3}}
\put(   11.00000,   6.40386){\circle*{0.3}}
\put(   12.20000,   9.19233){\circle*{0.3}}
\put(   13.40000,  10.55928){\circle*{0.3}}
\put(   14.60000,   9.19233){\circle*{0.3}}
\put(   15.80000,   6.40386){\circle*{0.3}}
\put(   17.00000,   4.25556){\circle*{0.3}}
\put(   18.20000,   3.31078){\circle*{0.3}}
\put(   19.40000,   3.05162){\circle*{0.3}}
\put(   20.60000,   3.00575){\circle*{0.3}}
\put(   21.80000,   3.00043){\circle*{0.3}}
\put(   23.00000,   3.00000){\circle*{0.3}}
\put(   24.20000,   3.00000){\circle*{0.3}}
\put(   25.40000,   3.00000){\circle*{0.3}}
\put(   26.60000,   3.00000){\circle*{0.3}}
\put(   27.80000,   3.00000){\circle*{0.3}}
\put(   29.00000,   3.00000){\circle*{0.3}}
\put(    5.00000,3){\line(0,1){    .00043}}
\put(    6.20000,3){\line(0,1){    .00575}}
\put(    7.40000,3){\line(0,1){    .05162}}
\put(    8.60000,3){\line(0,1){    .31078}}
\put(    9.80000,3){\line(0,1){   1.25556}}
\put(   11.00000,3){\line(0,1){   3.40386}}
\put(   12.20000,3){\line(0,1){   6.19233}}
\put(   13.40000,3){\line(0,1){   7.55928}}
\put(   14.60000,3){\line(0,1){   6.19233}}
\put(   15.80000,3){\line(0,1){   3.40386}}
\put(   17.00000,3){\line(0,1){   1.25556}}
\put(   18.20000,3){\line(0,1){    .31078}}
\put(   19.40000,3){\line(0,1){    .05162}}
\put(   20.60000,3){\line(0,1){    .00575}}
\put(   21.80000,3){\line(0,1){    .00043}}
\put(   23.00000,3){\line(0,1){    .00000}}
\put(   24.20000,3){\line(0,1){    .00000}}
\put(   25.40000,3){\line(0,1){    .00000}}
\put(   26.60000,3){\line(0,1){    .00000}}
\put(   27.80000,3){\line(0,1){    .00000}}
\put(   29.00000,3){\line(0,1){    .00000}}
\put(   40.00000,   3.00000){\circle*{0.3}}
\put(   41.20000,   3.00000){\circle*{0.3}}
\put(   42.40000,   3.00000){\circle*{0.3}}
\put(   43.60000,   3.00000){\circle*{0.3}}
\put(   44.80000,   3.00000){\circle*{0.3}}
\put(   46.00000,   3.00000){\circle*{0.3}}
\put(   47.20000,   3.00000){\circle*{0.3}}
\put(   48.40000,   3.00000){\circle*{0.3}}
\put(   49.60000,   3.00000){\circle*{0.3}}
\put(   50.80000,   3.00000){\circle*{0.3}}
\put(   52.00000,   3.00863){\circle*{0.3}}
\put(   53.20000,   3.19970){\circle*{0.3}}
\put(   54.40000,   4.88334){\circle*{0.3}}
\put(   55.60000,  10.23870){\circle*{0.3}}
\put(   56.80000,  14.33892){\circle*{0.3}}
\put(   58.00000,  10.23870){\circle*{0.3}}
\put(   59.20000,   4.88334){\circle*{0.3}}
\put(   60.40000,   3.19970){\circle*{0.3}}
\put(   61.60000,   3.00863){\circle*{0.3}}
\put(   62.80000,   3.00000){\circle*{0.3}}
\put(   64.00000,   3.00000){\circle*{0.3}}
\put(   40.00000,3){\line(0,1){    .00000}}
\put(   41.20000,3){\line(0,1){    .00000}}
\put(   42.40000,3){\line(0,1){    .00000}}
\put(   43.60000,3){\line(0,1){    .00000}}
\put(   44.80000,3){\line(0,1){    .00000}}
\put(   46.00000,3){\line(0,1){    .00000}}
\put(   47.20000,3){\line(0,1){    .00000}}
\put(   48.40000,3){\line(0,1){    .00000}}
\put(   49.60000,3){\line(0,1){    .00000}}
\put(   50.80000,3){\line(0,1){    .00000}}
\put(   52.00000,3){\line(0,1){    .00863}}
\put(   53.20000,3){\line(0,1){    .19970}}
\put(   54.40000,3){\line(0,1){   1.88334}}
\put(   55.60000,3){\line(0,1){   7.23870}}
\put(   56.80000,3){\line(0,1){  11.33892}}
\put(   58.00000,3){\line(0,1){   7.23870}}
\put(   59.20000,3){\line(0,1){   1.88334}}
\put(   60.40000,3){\line(0,1){    .19970}}
\put(   61.60000,3){\line(0,1){    .00863}}
\put(   62.80000,3){\line(0,1){    .00000}}
\put(   64.00000,3){\line(0,1){    .00000}}

\put(4.2,1.8){$\scriptscriptstyle{0}$}
\put(28.5,1.8){$\scriptscriptstyle{20}$}
\put(16.5,1.8){$\scriptscriptstyle{10}$}
\put(10.5,1.8){$\scriptscriptstyle{5}$}
\put(22.5,1.8){$\scriptscriptstyle{15}$}

\put(39.2,1.8){$\scriptscriptstyle{0}$}
\put(63.5,1.8){$\scriptscriptstyle{20}$}
\put(51.5,1.8){$\scriptscriptstyle{10}$}
\put(45.5,1.8){$\scriptscriptstyle{5}$}
\put(57.5,1.8){$\scriptscriptstyle{15}$}

\put(31,3.3){$\scriptstyle{n}$}
\put(66,3.3){$\scriptstyle{k}$}

\put(3,8.5){$\scriptscriptstyle{0.2}$}
\put(4.9, 9){\line(1,0){0.2}}
\put(3,14.5){$\scriptscriptstyle{0.4}$}
\put(4.9, 15){\line(1,0){0.2}}
\put(38,8.5){$\scriptscriptstyle{0.2}$}
\put(39.9, 9){\line(1,0){0.2}}
\put(38,14.5){$\scriptscriptstyle{0.4}$}
\put(39.9, 15){\line(1,0){0.2}}

\put(5.4,17){$\scriptscriptstyle{|\Psi (n)|^2}$}
\put(40.4,17){$\scriptscriptstyle{|\mathcal{F}[\Psi ](k)|^2}$}

\end{picture}
\caption{$|\Psi (n)|^2$ and $|\mathcal{F}[\Psi ](k)|^2$ in the case when $N\!=\!21$, $\kappa \!=\!\frac{2}{3}$,  $n_0\!=\!7$ and  $k_0\!=\!14$.}
\end{figure}
\end{center}
For any $n_0,\, k_0\in \mathbb{Z}_N=\{0,1,..., N\!-\!1\}$, the Fourier transform of the function
\[
\Psi  :\{0,1,..., N\!-\!1\}\longrightarrow \mathbb{C}, \qquad \Psi (n)={\rm e}^{\frac{2\pi {\rm i}}{N}k_0n}\, \Upsilon _\kappa (n\!-\!n_0)
\]
is 
\[
\mathcal{F}[\Psi ](k)={\rm e}^{\frac{2\pi {\rm i}}{N}(k_0-k
)n_0}\, \Upsilon _{\frac{1}{\kappa }}(k\!-\!k_0).
\]
Particularly, we have (see Figure 2)
\[
| \Psi (n)|^2=|\Upsilon _\kappa (n\!-\!n_0)|^2\qquad {\rm and} \qquad |\mathcal{F}[\Psi ](k)|^2=| \Upsilon _{\frac{1}{\kappa }}(k\!-\!k_0)|^2.
\]

\section{A quantum-like description for price-ownership}

We consider a stock market with a large number $N$ of traders and investigate the price of a fixed stock. We use a unit of cash chosen such that the stock price  is less than $N$ units at any moment of time. Assuming that the only possible values of the price of the considered stock are $0$, $1$, $2$, ..., $N\!-\!1$,  we describe the stock price at a fixed moment of time by using a normalized function
\[
\Phi  :\{ 0,1,2,...,N\!-\!1\}\longrightarrow \mathbb{C}
\]
chosen such that $|\Phi (n)|^2$ represents the probability to have a price equal to $n$ units of cash if a transaction takes place. We label the traders (in a more or less natural way) by using $T_0$, $T_1$,  ... ,$T_{N-1}$ and consider that the number $|F[\Phi ](k)|^2$ represents the probability that the stock owner is $T_k$.

In the case when the stock price is described by  $\delta _m$  we have (see Figure 1)
\[
|\delta _m(n)|^2=\left\{
\begin{array}{lll}
1 & {\rm if} & n=m\\
0 & {\rm if} & n\neq m
\end{array} \right.\qquad {\rm and}\qquad |\mathcal{F}[\delta _m ](k)|^2=\frac{1}{N}.
\]
This means that the stock price is $m$ (units of cash), and the probability for each trader to be the owner is the same, namely, $1/N$. We know with precision the price but we have no information on the stock owner.

In the case when the stock price is described by  the function $\Psi$  from Example 2, we have only partial information on the price, but we have at the same time partial information on the owner.  The stock owner is $T_0$, $T_1$,  ... ,$T_{N-1}$ with the probabilities $|\mathcal{F}[\Psi ](0)|^2$,  $|\mathcal{F}[\Psi ](1)|^2$, ... , $|\mathcal{F}[\Psi ](N\!-\!1)|^2$, respectively (see Figure 2). If a transaction  takes place, the stock price is $0$, $1$, ..., $N\!-\!1$ with the probabilities $|\Psi (0)|^2$,  $|\Psi (1)|^2$, ... , $|\Psi (N\!-\!1)|^2$, respectively.

The functions $\delta _0$,  $\delta _1$, ... , $\delta _{N-1}$ form an orthonormal  basis in the space $\mathcal{H}$,
\[
\langle \delta _n,\delta _m\rangle =\left\{
\begin{array}{lll}
1 & {\rm if} & n=m\\
0 & {\rm if} & n\neq m
\end{array} \right.
\]
 and  any function $\Phi $ describing the stock price is a superposition of these functions
\[
\Phi =\sum _{m=0}^{N-1}\Phi (m)\, \delta _m\qquad {\rm with}\qquad \Phi (m)=\langle \delta _m,\Phi \rangle.
\]

\section{Price and ownership operators}

In quantum mechanics, in the case of a particle moving along an axis, the position and momentum are described by the linear operators $\psi \mapsto \hat x \, \psi $ and $\psi \mapsto \hat p \, \psi $ \, defined as
\[
(\hat x \, \psi )(x)=x\, \psi (x)\qquad {\rm and}\qquad \hat p \, \psi =-{\rm i}\hbar \frac{d\psi }{dx}.
\]
They satisfy the relations
\[
\hat p=F^{-1}\hat xF\qquad {\rm and}\qquad [\hat x,\hat p]={\rm i}\hbar 
\]
where $[\hat x,\hat p]\!=\!\hat x \hat p\!-\!\hat p\hat x$ is the commutator of $\hat x$,  $\hat p$,
and $\hbar $ is Planck constant divided by $2\pi $. 

The mapping 
\[
\wp :\mathcal{H}\longrightarrow \mathcal{H}:\varphi \mapsto \wp\, \varphi ,\qquad (\wp\, \varphi )(n)=n\, \varphi (n)
\]
is a linear operator, we call the {\it price operator}. Since 
\[
\wp \, \delta _m=m\, \delta _m
\]
the function $\delta _m$ is an eigenfunction of $\wp $
corresponding to the eigenvalue $m$. 
The eigenfunctions of $\wp $ are exactly the functions corresponding to the cases when the stock price has an exactly known value. If $\Phi $ is a normalized function then the number
\[
\langle \wp \rangle =\langle \Phi ,\wp \Phi \rangle =\sum _{n=0}^{N-1}n\, |\Phi (n)|^2
\]
represent the {\it mean value} of the  price.

The linear operator defined in terms of the finite Fourier transform as
\[
\mathcal{O} :\mathcal{H}\longrightarrow \mathcal{H}, \qquad \mathcal{O}=\mathcal{F}^{-1}\wp \mathcal{F}
\]
represents the {\it ownership operator}. Since 
\[
\mathcal{O} \, (\mathcal{F}^{-1}[\delta _m])=m\,(\mathcal{F}^{-1}[\delta _m])
\]
the function $\mathcal{F}^{-1}[\delta _m]$ is an eigenfunction of $\mathcal{O} $
corresponding to the eigenvalue $m$, and 
\[
\Phi =\sum_{m=0}^{N-1}a_m\, \mathcal{F}^{-1}[\delta _m]\quad \Longrightarrow \quad
a_m=\langle \mathcal{F}^{-1}[\delta _m],\Phi \rangle =\langle \delta _m, \mathcal{F}[\Phi ]\rangle =\mathcal{F}[\Phi ](m).
\]

The operators $\wp $ and $\mathcal{O}$ play a role similar to $\hat x$ and $\hat p$. 
If $N$ is large enough, then
\[
[\wp ,\mathcal{O}]\approx (N/2\pi )\, {\rm i}.
\]
For example (see Table 1), in the case $N=21$ most of the eigenvalues of the operator $[\wp ,\mathcal{O}]=\wp  \mathcal{O}-\mathcal{O}\!  \wp $ are approximatively equal to $21/2\pi =3.3422538049298020511...$

In quantum mechanics \cite{Messiah} , two variables are said to be compatible if they can simultaneously be defined with infinite precision. Two compatible variables are represented by commuting linear operators. Since $\wp \mathcal{O}\neq \mathcal{O}\wp $, the variables $\wp $ and $\mathcal{O}$ are complementary observables which cannot be defined simultaneously with infinite accuracy.
\begin{table} 
\begin{tabular}{rrrrrr}
\hline
$n$ & $\lambda _n$\qquad \quad \ \mbox{} &   $n$ & $\lambda _n$\qquad \quad \ \mbox{}  & $n$ &  $\lambda _n$\qquad \quad \ \mbox{} \\ 
\hline
1 & -133.965206767811\, {\rm i} & 8 & 3.342253797136\, {\rm i} & 15 & 3.342253907182\, {\rm i}\\
2 & -27.116000868817\, {\rm i} &  9 & 3.342253804904\, {\rm i} & 16 & 3.342264884479\, {\rm i}\\
3 & 0.563232849284\, {\rm i} & 10 &  3.342253804929\, {\rm i} & 17 & 3.342954064008\, {\rm i}\\
4 & 3.198831527436\, {\rm i} & 11 & 3.342253804929\, {\rm i} & 18 & 3.369561581989\, {\rm i}\\
5 & 3.337619084687\, {\rm i} & 12 & 3.342253804929\, {\rm i}& 19 & 4.015171698810\, {\rm i}\\
6 & 3.342159991389\, {\rm i} & 13 & 3.342253804930\, {\rm i} & 20 & 13.739015316163\, {\rm i}\\
7 & 3.342252660619\, {\rm i} & 14 & 3.342253805426\, {\rm i} & 21 & 92.750113443389\, {\rm i}\\
\hline
\end{tabular}
\caption{The eigenvalues $\lambda _n$ of the commutator $[\wp ,\mathcal{O}]$ in the case $N=21$.}
\end{table}

\section{Limit imposed by uncertainty relation}

In quantum mechanics, the uncertainties in $\hat x$ and $\hat p$, described by using the root-mean-square deviations
$\Delta \hat x=\sqrt{\langle \hat x^2\rangle -\langle x\rangle ^2}$ and  $\Delta \hat p=\sqrt{\langle \hat p^2\rangle -\langle p\rangle ^2}$ satisfy the relation
\[
\Delta \hat x\cdot \Delta \hat p\geq \frac{1}{2}\hbar .
\]
From the used mathematical formalism it follows that the uncertainties in $\wp $ and $\mathcal{O}$
\[
\Delta \wp =\sqrt{\langle  \wp ^2\rangle -\langle \wp \rangle ^2}=\sqrt{\langle \Phi ,\wp ^2\Phi \rangle -\langle \Phi ,\wp \Phi \rangle ^2}
\]
\[
\Delta \mathcal{O} =\sqrt{\langle  \mathcal{O} ^2\rangle -\langle \mathcal{O} \rangle ^2}=\sqrt{\langle \Phi ,\mathcal{O} ^2\Phi \rangle -\langle \Phi ,\mathcal{O} \Phi \rangle ^2}
\]
satisfy the generalized uncertainty relation
\begin{equation}\label{uncertrel}
\Delta \wp \cdot \Delta \mathcal{O}\geq \frac{1}{2} |\langle [\wp ,\mathcal{O}]\rangle |
\end{equation}
where $\langle [\wp ,\mathcal{O}]\rangle =\langle \Phi ,[\wp ,\mathcal{O}]\Phi \rangle $.
Generally, the accuracy in $\wp $ cannot be improved without a corresponding loss in the accuracy of $\mathcal{O}$. In the case when $\Delta \wp \neq 0$, $\Delta \mathcal{O}\neq 0$ and
\[
\Delta \wp \cdot \Delta \mathcal{O}= \frac{1}{2} |\langle [\wp ,\mathcal{O}]\rangle |
\]
we have a maximum of information concerning both the price and the ownership.
For example, in the case of the Gaussian type function (see Example 2)
\[
\Psi  :\{0,1,..., 20\}\longrightarrow \mathbb{C}, \qquad
\Psi (n)={\rm e}^{\pi {\rm i}n}\, \Upsilon _1 (n\!-\!10)
\]
we have
$\Delta \wp \cdot \Delta \mathcal{O}\approx 1.6711269024646$  and $\frac{1}{2} \frac{21}{2\pi}\approx 1.6711269024649 $, that is, 
\[
\Delta \wp \cdot \Delta \mathcal{O}\approx  \frac{1}{2} |\langle [\wp ,\mathcal{O}]\rangle |.
\]
\section{On the time evolution of the stock price}

The main purpose of a mathematical model concerning the stock price is to anticipate the price.  We can only predict probabilities. By following the analogy with the quantum mechanics, we assume that the time dependent function $\Phi (n,t)$ describing the stock price satisfies a Schr\" odinger type equation
\[
{\rm i}\, \frac{\partial }{\partial t}\Phi =\left( \frac{\mathcal{O}^2}{2\mu }+\mathcal{V}(\wp ,t)\right)\Phi 
\]
where $\mu $ is a positive parameter and $\mathcal{V}(\wp ,t)$ is a time dependent function describing the interactions between traders as well as the external economic conditions. The function $\Phi (n,t)$ is well-determined if its values $\Phi (n,t_0)$ at a fixed moment of time $t_0$ are known.
A large variety of cases can be easily investigated by using, for example, the computer program in MATHEMATICA presented in \cite{Cotfas}.

\section{Concluding remarks}

The relation price-ownership is similar to the relation position-momentum from quantum mechanics. The proposed description, based on the mathematical formalism used in the the case of quantum systems with finite-dimensional Hilbert space, may help us to better understand this relation playing a fundamental role in finance.


\end{document}